\documentclass[reprint, amsmath, amssymb, aps, showkeys]{revtex4-2}

\usepackage{graphicx}
\usepackage[colorlinks=true, allcolors=blue]{hyperref}
\usepackage{multirow}
\usepackage{float}
\usepackage{mathtools} % http://ctan.org/pkg/mathtools
\usepackage{xcolor}

\newcommand{\hlf}{\frac{1}{2}}
\newcommand{\lt}{\left}
\newcommand{\rt}{\right}

\begin{document}

\preprint{APS/diffusion}

\title{Hamiltonian chaos for one particle with two waves: Self-consistent dynamics}

% region authors
\author{Matheus J. Lazarotto}
\email{matheus\_jean\_l@hotmail.com}
\affiliation{Aix-Marseille Universit\'{e}, CNRS, UMR 7345 PIIM, F-13397, Marseille cedex 13, France}
\affiliation{Instituto de F\'{i}sica, Universidade de S\~{a}o Paulo, Rua do Mat\~{a}o 1371, S\~{a}o Paulo 05508-090, Brazil}

\author{Iber\^{e} L. Caldas}
\email{ibere@if.usp.br}
\affiliation{Instituto de F\'{i}sica, Universidade de S\~{a}o Paulo, Rua do Mat\~{a}o 1371, S\~{a}o Paulo 05508-090, Brazil}

\author{Yves Elskens}
\email{yves.elskens@univ-amu.fr}
\affiliation{Aix-Marseille Universit\'{e}, CNRS, UMR 7345 PIIM, F-13397, Marseille cedex 13, France}
% endregion

\date{\today}

\begin{abstract}
    A simple model of wave-particle interaction is studied in its self-consistent form, that is, where 
    the particles are allowed to feedback on the waves dynamics.
    We focus on the configurations of locked solutions (equilibria) and how the energy-momentum exchange 
    mechanism induces chaos in the model. 
    As we explore the system, we analyse the mathematical structure that gives rise to locked states 
    and how the model's non-linearity enables multiple equilibrium amplitudes for waves.
    We also explain the predominance of regularity as we vary the control parameters and the mechanism 
    behind the emergence of chaos under limited parameter choices. 
\end{abstract}


\maketitle

\textbf{
When modeling physical systems composed of waves and particles, it is often assumed that the 
latter are subject to the influence of the former, but not the other way around. 
In this work, we investigate the effect of considering the mutual interaction between waves and 
particles, resulting in a self-consistent dynamics.
With a Hamiltonian model of the system, we consider the case of one particle interacting with 
two waves. 
We first reduce the equations from a 6-dimensional problem to a 4-dimensional one, with one constant 
of the motion, enabling better handling of equations and visualization of their solutions. 
While previous studies analyzed how a particle influenced by two free waves -- i.e. without 
mutual interaction -- behaves chaotically, here we discuss how the particle serves as a mediator, 
enabling the exchange of energy and momentum between the waves and propagating its chaotic motion 
through their evolution.
Moreover, we find that several locked states, where both waves travel in equilibrium with the particle, 
can exist for the same total momentum and energy. 
}

\section{Introduction}\label{sec:introduction}

% \textcolor{magenta}{Context intro (plasma physics):} 
When describing a plasma on a microscopic scale, a fundamental process is the interaction between its particles 
(electrons and ions) and waves propagating in the plasma~\cite{Yves_book, Escande_1}. 
In the particular limit of collisionless plasmas, that is when the time scale of particle `collisions' is 
much longer than the time scale of the interaction with waves, the energy and momentum exchange between 
particles and waves becomes relevant to the point of generating instabilities, favoring wave amplification 
or particle acceleration~\cite{Balescu}. 

% \textcolor{magenta}{Effects of wave-particle interaction:}
For instance, this scenario can be experimentally reproduced in traveling wave tubes (TWT), where a 
beam of electrons interacts with electrostatic waves in a controllable environment~\cite{Doveil_1, Sousa}. 
This setup enables the analysis of chaotic interactions due to the energy-momentum exchange that 
may lead to instabilities and further turbulence in plasmas, which in turn correlates to the 
emergence of chaotic regions in phase space. 
This chaotic transition can be caused by resonance overlap of waves, nonlinear synchronization 
by a single nonresonant wave or the `devil's staircase' mechanism, causing a spread of the velocity 
distribution of the beam~\cite{Doveil_2,Ott}.
Stability islands on the other hand can correlate with coherent particle acceleration (beam trapping), 
and the breaking of barriers (invariant tori in the one-particle phase space) allowing for new ways of chaos control. 
% -- Doveil and Macor: break of barriers of invariant tori promote chaos and acceleration of particles in a 
%                      devil's staricase model -- seen in the experiment for electrons as test particles  

% \textcolor{magenta}{Context theoretical model (Langmuir waves)}
When modeling such scenarios, the Langmuir waves, that is the collective vibration of electrons 
in presence of much heavier ions that neutralize the system, are found to be coupled with 
quasi-resonant particles (i.e.\ electron's velocity being close to waves phase velocity)\cite{Smith}. 
Thus, for simplicity, the usual Hamiltonian models initially consider only the effect of waves 
on the particles. 
To give a step ahead, the re-interaction of the particles with the waves can be taken into account, 
therewith making the dynamics self-consistent~\cite{Yves_book,Escande_1,Besse}. 
Consequently, electrons (and ions) do not play just the role of test particles but instead promote 
energy-momentum exchange with the waves, allowing for the aforementioned nonlinear phenomena. 
To model it, one can consider $M$ independent Langmuir waves behaving as harmonic oscillators and 
$N$ quasi-resonant free charged particles, with the addition of a coupling term for 
each wave-particle pair~\cite{ONeil,Onishchenko, Mynick}.

Gomes \textit{et al.}~\cite{Gomes} considered two of the simplest cases, namely that of one 
wave and one particle ($M=1,N=1$), which is analytically treatable and shows the electron either being 
trapped (strong resonance) or passing through wave potential wells~\cite{Adam,Castillo-Negrete}, and 
the case of one wave interacting with two particles ($M=1,N=2$), that presents chaotic behavior. 
We direct the reader to the work of Gomes \textit{et al.}~\cite{Gomes} for a simple yet broad introduction 
on the experimental context, modeling and applications of wave-particle dynamics.

% \textcolor{magenta}{Purpose:}
In this sense, our work is complementary to Gomes' work, for we consider the remaining case of two waves 
interacting with one particle ($M=2,N=1$), aiming to a simple description of the interaction between 
waves and particles. 
By studying these simplest cases, one may shed light on the microscopic mechanisms that induce chaos 
in the macroscopic system and how to prevent (or allow) it. 
In the current work, for instance, it is shown that equilibrium solutions only come in locked configurations 
allowing for multiple unstable wave amplitudes combinations.
The emergence of chaos in the system is predominantly due to separatrix chaos related to the parameter limit 
where the non-linear coupling produces energy-momentum exchange.

% \textcolor{magenta}{Paper outline:} 
In what follows, Sec.~\ref{sec:model-generic} starts by presenting the generic wave-particle 
self-consistent Hamiltonian model and Sec.~\ref{sec:model-reduction} the particular case studied 
here (for $M=2, N=1$), along with its mathematical simplification to reduce the number of degrees 
of freedom and number of parameters. 
Sec.~\ref{sec:equilibrium-solutions} presents a global analysis of the equilibrium solutions 
in parameter space.
Sec.~\ref{sec:phase-space} discusses the emergence of chaos in phase space and the integrable 
limits of the system.
Appendix sections provide more details on the model parameter simplifications, 
the stability of equilibrium solutions, and the decoupling limit of the system equations of motion. 

\section{The wave-particle Hamiltonian}\label{sec:model}
\subsection{The generic model}\label{sec:model-generic}

Within the context of Langmuir waves interacting with ions in plasmas, one can model the self-consistent 
dynamics of $N$ identical particles traveling in a periodically bounded interval of length $L$ via 
a coupling with $M$ longitudinal waves. 
Each particle is described by generalized coordinate-momentum pairs ($x_i, p_i$) and each wave written 
in phasor formulation as $Z_j = X_j + {\rm i} Y_j = \sqrt{2 I_j}\,\mathrm{e}^{-{\rm i} \theta_j}$, with either 
cartesian ($X_j, Y_j$) or polar ($\theta_j, I_j$) canonical variables. 
Within this interval, wave numbers are given by $k_j = \frac{2 \pi j }{L}$, for $j \in \mathbb{Z}$, and 
natural frequencies by $\omega_{0j}$.
The system's Hamiltonian is then~\cite{Yves_book,Escande_1}
\begin{equation}\label{eq:hamiltonian-full-carte}
\begin{split}
    H_{\textrm{sc}}^{N,M} = &\sum_{i=1}^N \frac{p_i^2}{2m} + 
                             \sum_{j=1}^M \omega_{0j} \frac{X_j^2 + Y_j^2}{2} \\ 
                            &+\epsilon \sum_{i=1}^N \sum_{j=1}^M \frac{\beta_j}{k_j}\lt(Y_j\sin(k_j x_i) - X_j\cos(k_j x_i)\rt),
\end{split}
\end{equation}
or equivalently in polar-phasor coordinates
\begin{equation}\label{eq:hamiltonian-full-polar}
\begin{split}
    H_{\textrm{sc}}^{N,M} = &\sum_{i=1}^N \frac{p_i^2}{2m} + 
                             \sum_{j=1}^M \omega_{0j} I_j \\
                            &-\epsilon \sum_{i=1}^N \sum_{j=1}^M \frac{\beta_j}{k_j} \sqrt{2 I_j}\cos(k_j x_i - \theta_j),
\end{split}
\end{equation}
with $\beta_j$ as a coupling constant with the $j$th-wave and $\epsilon$ an overall coupling scale. 
For the wave degrees of freedom, $X_j$ and $\theta_j$ are treated as coordinates with $Y_j$ and $I_j$ 
as their respective conjugate momenta. 

The Hamiltonian $H_{\textrm{sc}}^{N,M}$ comprises three contributions: the free motion (kinetic energy) 
of particles ($m > 0$); the harmonic oscillation of waves ($\omega_{0j} > 0$); and the coupling between 
particles and waves.
Furthermore, the Hamiltonian $H_{\textrm{sc}}^{N,M}$ is invariant under translations in time and in space, so that the total energy $E = H_{\textrm{sc}}^{N,M}$ and the total momentum 
$P = \sum_{i=1}^N p_i + \sum_{j=1}^M k_j I_j$ are conserved. 
The latter constant reveals that the growth or decay of a wave with positive phase velocity ($\frac{\omega_{0j}}{k_j} > 0$) 
is directly balanced with the slowing down or acceleration of particles. 
On interchanging the time and space variables, this Hamiltonian also captures the key physics in the 
dynamics of traveling wave tubes~\cite{Escande_1,Doveil_1}.
For instance, in the paradigmatic two-wave scenario with a slaved particle, given parameters 
$\epsilon \beta_1 = \epsilon \beta_2 = \beta, k_1 = k_2 = k, m$, the Chirikov overlap parameter~\cite{Chirikov_2, Escande_2} 
is $s = 2 \sqrt{\beta k / m} ( {(2 I_1)}^{1/4} + {(2 I_2)}^{1/4} ) / | \omega_1 - \omega_2 |$.

\subsection{The single particle with two waves}\label{sec:model-reduction}

To address the key physics of the system, one can start looking at the simpler scenarios, namely 
those with few waves and particles interacting with each other. 
The simplest case, that of $M=N=1$, is integrable and was treated before in~\cite{Adam,Castillo-Negrete,Gomes}; 
thanks to its integrability, it provides reference information 
for the more complex cases where chaos emerges.
With the addition of a particle ($N=2$, $M=1$), as studied by Gomes \textit{et al.}~\cite{Gomes}, 
the non-linear term in Hamiltonian (\ref{eq:hamiltonian-full-carte}) implies the emergence of 
chaos in phase space not only around separatrices but also near elliptic points. 

Here, we consider the `mirror' case of one particle coupled to two waves, that is the case with $N=1$, 
$M=2$ in Hamiltonian (\ref{eq:hamiltonian-full-carte}) (\ref{eq:hamiltonian-full-polar}). 
The system thus has 7 free parameters: $k_1, k_2, \epsilon\beta_1, \epsilon\beta_2, \omega_{01}, 
\omega_{02}$ and $m$ (along with 3 free scales, leaving 4 dimensionless free parameters). 
For simplicity, the $\epsilon\beta_j$ are selected to be equal and unity: $\epsilon\beta_1 = 
\epsilon\beta_2 = 1$, as well as the particle mass: $m=1$, which can be set without loss of generality 
by rescaling variables (see Appendix~\ref{sec:append:param-reduc} for more details).

Given the translational invariance, only the relative positions $\theta_j - k_j x$ are dynamically 
relevant; 
thus, for commensurate $k_1$ and $k_2$, the particle position $x$ can meet periodic boundary conditions, 
$x \in [0, L]$, compatible with both waves. For incommensurate wave numbers, the position space is 
necessarily the full real line. 
By setting equal wave numbers and given a proper rescaling of space, we select $k_1 = k_2 = 1$ 
(in order to keep free phase-velocities, the frequencies $\omega_{0j}$ are kept different).

The resulting 3-degrees-of-freedom Hamiltonian is autonomous and translation-invariant; 
in cartesian-phasor coordinates it reads
\begin{equation}\label{eq:hamiltonian-n2m1-carte}
\begin{split}
    H^{1,2}_\textrm{sc} =&  \frac{p^2}{2} 
                          + \omega_{01} \frac{X_1^2 + Y_1^2}{2} 
                          + \omega_{02} \frac{X_2^2 + Y_2^2}{2} \\
                         &+ (Y_1 + Y_2) \sin(x) - (X_1 + X_2) \cos(x), \\
\end{split}
\end{equation}
and likewise in polar-phasor form
\begin{equation}\label{eq:hamiltonian-n2m1-polar}
\begin{split}
    H^{1,2}_\textrm{sc} =& \frac{p^2}{2} 
                          + \omega_{01} I_1 + \omega_{02} I_2 \\
                         &- \sqrt{2 I_1}\cos(x - \theta_1) 
                          - \sqrt{2 I_2}\cos(x - \theta_2).
\end{split}
\end{equation}

To further simplify the model and better visualize solutions in phase space, 
it is convenient to reduce the number of degrees of freedom, which is obtained 
with a simple sequence of canonical trasformations.
Although assuming equal wave-lengths and coupling strengths, the transformations 
below do not depend on these assumptions.

The particle's degree of freedom can be suppressed with the aid of the total momentum 
constant $P$ using $p = P - I_1 - I_2$. 
Besides, a Galilean transformation to a reference frame traveling with the first wave, 
provided by the generating function 
\begin{equation*}
\begin{split}
    G(x, \theta_1, \theta_2, \bar{p}, \bar{I}_1, \bar{I_2}) &=  (x - \omega_{01}t)(\bar{p} + \omega_{01}) 
                                                                + (\theta_1 - \omega_{01}t) \bar{I}_1 \\
                                                         &\quad + (\theta_2 - \omega_{01}t) \bar{I}_2 
                                                                + \frac{\omega_{01}^2 t}{2},
\end{split}
\end{equation*}
and the definition of the relative angle $\phi_j \coloneqq \bar{\theta}_j - \bar{x} = 
\theta_j - x$, provided by the generating function
\begin{equation*}
\begin{split}
    F(\bar{x}, \bar{\theta_1}, \bar{\theta}_2, p', I'_1, I'_2) &= I'_1 (\bar{\theta}_1 - \bar{x}) 
                                                                + I'_2 (\bar{\theta}_2 - \bar{x}) 
                                                                + p' \bar{x},
\end{split}
\end{equation*}
simplify Hamiltonian (\ref{eq:hamiltonian-n2m1-polar}) to
\begin{equation}\label{eq:hamiltonian-reduc-polar}
\begin{split}
    H' =&  \frac{\lt(I'_1 + I'_2\rt)^2}{2} 
         - \bar{P} \lt(I'_1 + I'_2\rt) 
         + \Delta_\omega I'_2 \\
        &- \sqrt{2 I'_1} \cos(\phi_1) 
         - \sqrt{2 I'_2} \cos(\phi_2),
\end{split}
\end{equation}
where prime symbols denote the final transformed variables 
$(x, \theta_j, p, I_j) 
\; \stackrel{G}{\mapsto} \;
(\bar{x}, \bar{\theta}_j, \bar{p}, \bar{I}_j) 
\; \stackrel{F}{\mapsto} \;
(x', \phi_j, p', I'_j)$
% \begin{equation*}
    % (x, \theta_j, p, I_j) 
    % \; \xrightarrow{G} \;
    % (\bar{x}, \bar{\theta}_j, \bar{p}, \bar{I}_j) 
    % \; \xrightarrow{F} \;
    % (x', \phi_j, p', I'_j),
% \end{equation*}
and will be dropped from now on.

From old to new variables, the amplitudes are unmodified ($I'_j = I_j$) and the new angles are the 
relative position between particle and wave $\phi_j = \theta_j - x$. The new total momentum is now 
$\bar{P} = P - \omega_{01}$ and we define the detuning parameter between the waves: $\Delta_\omega 
\coloneqq \omega_2 - \omega_1$, eliminating one of the frequencies. This final form has 2 degrees 
of freedom ($\phi_1, \phi_2$), their associated momenta ($I_1, I_2$), and 3 control 
parameters ($H, \bar{P}, \Delta_\omega$); the Hamiltonian $H$ itself being a constant of the motion. 
The frequency $\omega_{01}$ can be set to zero without loss of generality.

In cartesian variables, the Hamiltonian (\ref{eq:hamiltonian-n2m1-carte}) in simplified form reads
\begin{equation}\label{eq:hamiltonian-reduc-carte}
\begin{split}
    H' =& \frac{1}{8} \lt(u_1^2 + v_1^2 + u_2^2 + v_2^2\rt)^2 
        - \frac{\bar{P}}{2} \lt(u_1^2 + v_1^2 + u_2^2 + v_2^2\rt) \\
        &+ \frac{\Delta_\omega}{2} \lt(u_2^2 + v_2^2\rt)
         - u_1 - u_2.
\end{split}
\end{equation}
where 
\begin{equation}\label{eq:phasor-carte-transform}
    u_i = \sqrt{2 I_i} \cos\lt(\phi_i\rt), \quad v_i = \sqrt{2 I_i} \sin\lt(\phi_i\rt), 
\end{equation}
for $i = 1,2$.

It may be of interest to keep both forms since cartesian coordinates provide a smooth transition for 
the limit $I_j\to0$, though the polar ones provide a more intuitive picture of the dynamics. 
Appendix \ref{sec:append:cartesian-equations} shows the equations of motion in cartesian form.

It is worth mentioning that Hamiltonian (\ref{eq:hamiltonian-reduc-carte}) is fourth degree in both 
$u_i$ and $v_i$, making the visualization of phase space via Poincar\'{e} sections less straightforward.
Since the intersection of trajectories with the section may yield two positive (or negative) roots, 
the usual method of fixing a position (say $u_2=0$) and a sign condition on the conjugate momentum 
(say $v_2 > 0$), is not enough to produce a uniquely defined map. 
Since extra roots may exist and appear over the points of map made by the first set of roots, invariant 
circles appear overlapping. 
A solution can be achieved by filtering these two sets of roots with an extra condition as $\dot{u}_2 > 0$.

\section{Locked solutions}\label{sec:equilibrium-solutions}

As a reference for the system's global dynamics, one can look at the equilibrium 
solutions (i.e. $(\dot{I}_1, \dot{I}_2, \dot{\phi}_1, \dot{\phi}_2) = \vec{0}$) 
of the equations of motion provided by Hamiltonian (\ref{eq:hamiltonian-reduc-polar})
\begin{equation}\label{eq:motion-equations}
\begin{split}
    \dot{I}_1    &= -\partial_{\phi_1} H' = -\sqrt{2 I_1} \sin(\phi_1), \\ 
    \dot{I}_2    &= -\partial_{\phi_2} H' = -\sqrt{2 I_2} \sin(\phi_2), \\ 
    \dot{\phi}_1 &=  \partial_{I_1} H'    = (I_1 + I_2) - \bar{P} - \frac{\cos(\phi_1)}{\sqrt{2 I_1}}, \\ 
    \dot{\phi}_2 &=  \partial_{I_2} H'    = (I_1 + I_2) - \bar{P} + \Delta_\omega - \frac{\cos(\phi_2)}{\sqrt{2 I_2}}.
\end{split}
\end{equation}

In the present context, equilibrium solutions are also referred to as locked states, given that, when in 
equilibrium, the amplitude equations $(\dot{I}_1, \dot{I}_2)$ yield the condition that the relative 
phases are constant ($\phi_i^* \in \{0, \pi\}$, for $i=1,2$) meaning that both waves and the particle 
travel at the same velocity with the latter constrained to be at either the minimum or maximum of one 
of the waves while the other is in phase or anti-phase. 
It is worth mentioning that the last two equations in (\ref*{eq:motion-equations}) %($\dot{\phi}_1, \dot{\phi}_2$) 
prevent the existence of zero-wave solutions ($I_1(t)=I_2(t)=0$) for both waves simultaneously for 
generic ($\bar{P}, \Delta_\omega$). 
For single null waves, i.e. $I_1^* = 0 \neq I_2^*$ (or vice-versa), 
it is possible to obtain such solutions for $\Delta_\omega = 0$. 
In such case, the null wave has indefinite phase $\phi_i$ and the remaining one 
must have fixed amplitude given by the total momentum $\bar{P}$.
Other considerations on the $\Delta_\omega = 0$ case are discussed in Appendix~\ref{sec:append:null-detune}.

From the locked relative phases $\phi_1^* = n_1 \pi$ and $\phi_2^* = n_2 \pi$, for 
$n_i = 0, 1$ and $i=1,2$, the equilibrium amplitudes $I_1^*$ and $I_2^*$ thus must 
satisfy 
\begin{equation}\label{eq:equilibrium_I1}
    \frac{\cos(n_2 \pi)}{\sqrt{2 I_2^*}} - \frac{\cos(n_1 \pi)}{\sqrt{2 I_1^*}} = \Delta_\omega.
\end{equation}

% \begin{equation}\label{eq:equilibrium_I1}
    % I_1^* = \frac{I_2^*}{\lt((-1)^{n_2} - \Delta_\omega \sqrt{2I_2^*}\rt)^2}
% \end{equation}
% Conversely, the symmetrical expression for $I_1^* \to I_2^*$ is obtained by exchanging 
% $\Delta_\omega \to -\Delta_\omega$ and $n_2 \to n_1$. 
% One may note from eq.~\ref{eq:equilibrium_I1} that the phase $n_1$ is irrelevant for the 
% amplitude value $I_1^*$, although it will be relevant for its stability, as discussed ahead.

Therewith, the relation between amplitudes enables one to find ($I_1^*, I_2^*$) by substituting 
eq. (\ref{eq:equilibrium_I1}) into $\dot{I_1} = 0$, yielding
\begin{equation}\label{eq:equilibrium_surface}
\begin{split}
    R(I_2, n_2) & = I_2 \sqrt{2I_2} \lt(1\!+\!\lt((-1)^{n_2} - \Delta_\omega\sqrt{2I_2}\rt)^2 \rt) + \\
                &\!\! \lt(\lt(\Delta_\omega - \bar{P}\rt) \sqrt{2 I_2} - (-1)^{n_2} \rt) \lt((-1)^{n_2} - \Delta_\omega\sqrt{2I_2}\rt)^2.
\end{split}
\end{equation}

\begin{figure}[H]
    \centering
    \includegraphics[width=0.5\textwidth]{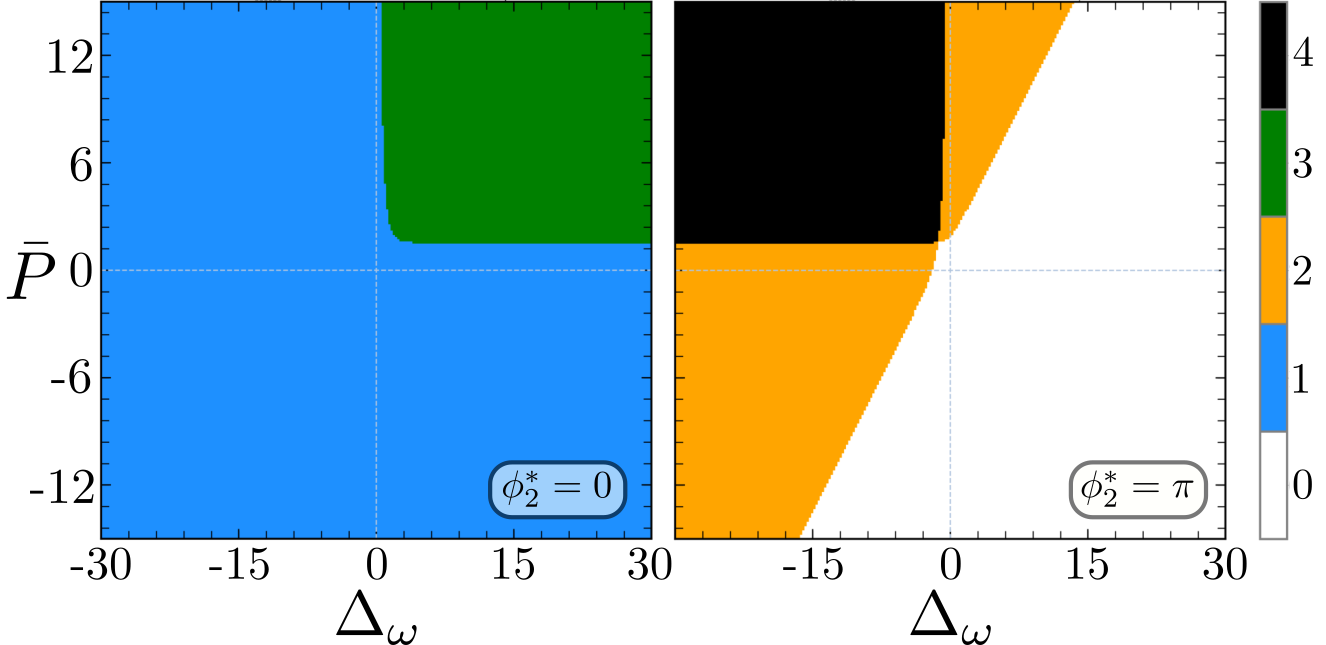}
    \caption{Color map with the number of equilibrium solutions for different $n_2$. 
             (Left) case $\phi_2^* = 0$; 
             (Right) case $\phi_2^* = \pi$. }
\label{fig:equilibrium_cmap}
\end{figure}

Solutions are then given by the condition $R(I_2^*, n_2) = 0$, with each one  
amounting for two locked states $(I_1^*, I_2^*, n_1 \pi, n_2 \pi)$, namely one 
with $n_1=0$ and the other with $n_1=1$. 
The existence of roots to condition (\ref{eq:equilibrium_surface}) was evaluated 
numerically by scanning over the parameter space $(\bar{P}, \Delta_\omega)$, with 
the result shown in figure~\ref{fig:equilibrium_cmap} for both cases $n_2 = 0$ and $n_2 = 1$.
 
\begin{figure}[H]
    \centering
    \includegraphics[width=0.40\textwidth]{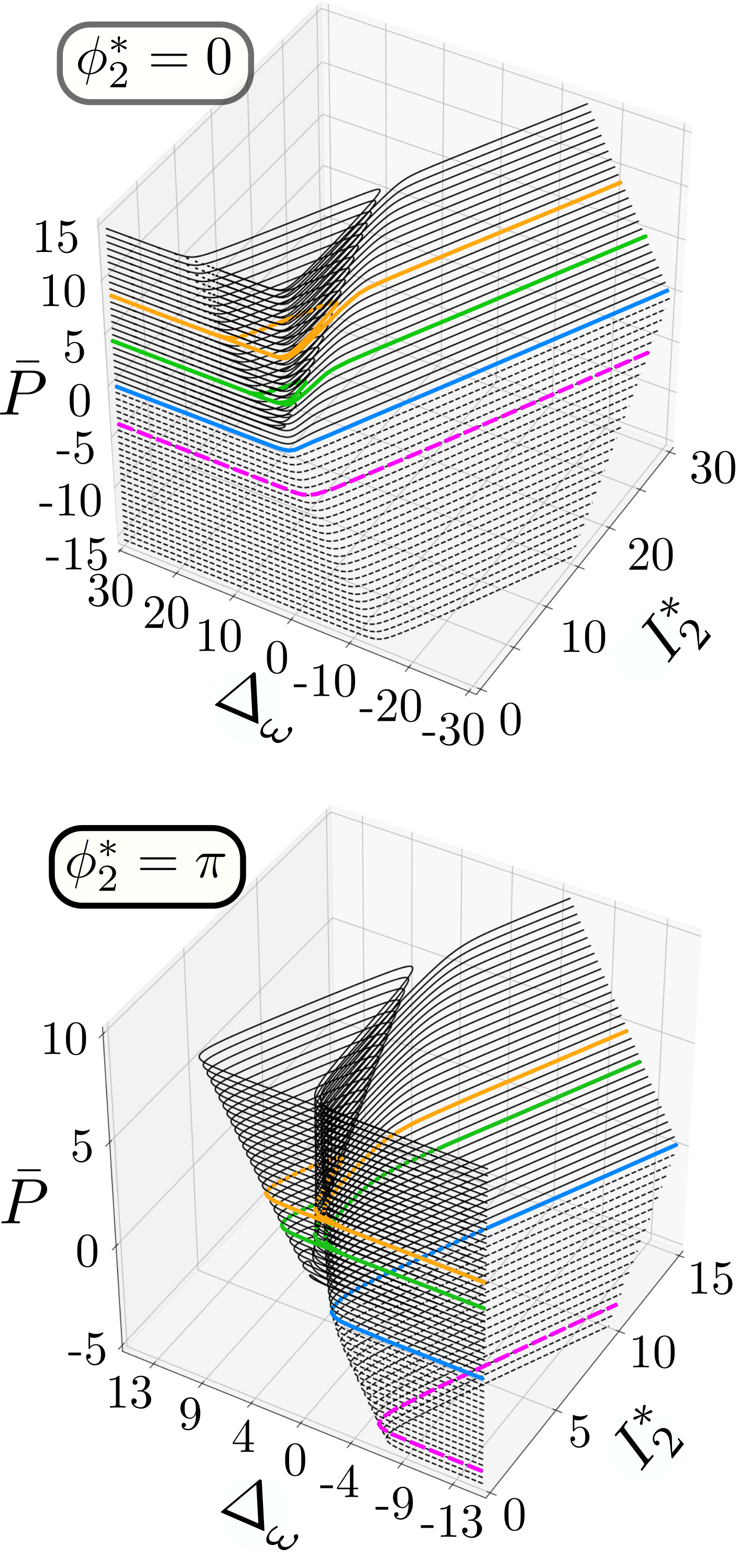}
    \caption{Surface plot of $R(I_2^*, \bar{P}, \Delta_\omega) = 0$ with its contour lines at different 
             constant values of $\bar{P}$.
             Dashed (resp.\ plain) lines correspond to $\bar{P} < 0$ (resp.\ $\bar{P} > 0$). 
             (Left) $\phi_2^* = 0$ (viz.\ $n_2 = 0$);
             (Right) $\phi_2^* = \pi$ (viz.\ $n_2 = \pi$). 
             Colored lines indicate particular values of $\bar{P}$ for the stability 
             analysis shown in figures \ref{fig:bifurcation_map_n1=0_n2=1} to 
             \ref{fig:bifurcation_map_n1=1_n2=1}.}
\label{fig:equilibrium_wireframe}
\end{figure}

In order to clarify the dependence of the number of roots as the control parameters change, 
figure~\ref{fig:equilibrium_wireframe} shows the surface $R(I_2^*) = 0$ along with the 
$I_2^*$ axis. In this plot, roots will exist whenever the $I_2^*$ axis crosses the surface for fixed 
control parameters $\bar{P}$ and $\Delta_\omega$.
 
For either case of $\phi_2^*$, the set of solutions is composed of two non-intersecting surfaces. 
One of them has a cup-like form, which asymptotically converges to $I_2^* \to 0$ as 
$\Delta_\omega \to \infty (-\infty)$ for $\phi_2^* = 0$ ($\phi_2^* = \pi$).
The second surface branch, completely detached from the cup, presents different concavity 
in each scenario of $\phi_2^*$. When $\phi_2^* = 0$, the branch is convex and therefore allows for 
one crossing throughout all parameter space, whereas for $\phi_2^* = \pi$ the concave surface allows 
only for crossings up to the limit $\Delta_\omega \lesssim 0$. 
The appearance of 2 extra roots, in both cases, is due to the cup-like branch, since any intersection 
with it implies two extra solutions, although these are only possible for $\bar{P} \gtrsim 1$. 
It is worth pointing out that the reasoning above does not consider any constraint on the Hamiltonian 
$H$ value.

From both figures~\ref{fig:equilibrium_cmap} and~\ref{fig:equilibrium_wireframe}, the case 
$\phi_2^* = 0$ guarantees that it is always possible that the particle is placed at the minimum of 
one of the waves and that it is locked to travel at the same speed, regardless of $\bar{P}, \Delta_\omega$. 
Moreover, if the total momentum of the system is large enough ($\bar{P}~\gtrsim~1$), new locked states 
become possible in which the waves may have enhanced or suppressed amplitude to provide balance of 
forces acting on the particle.

\begin{figure}[h!]
    \centering
    \includegraphics[width=0.50\textwidth]{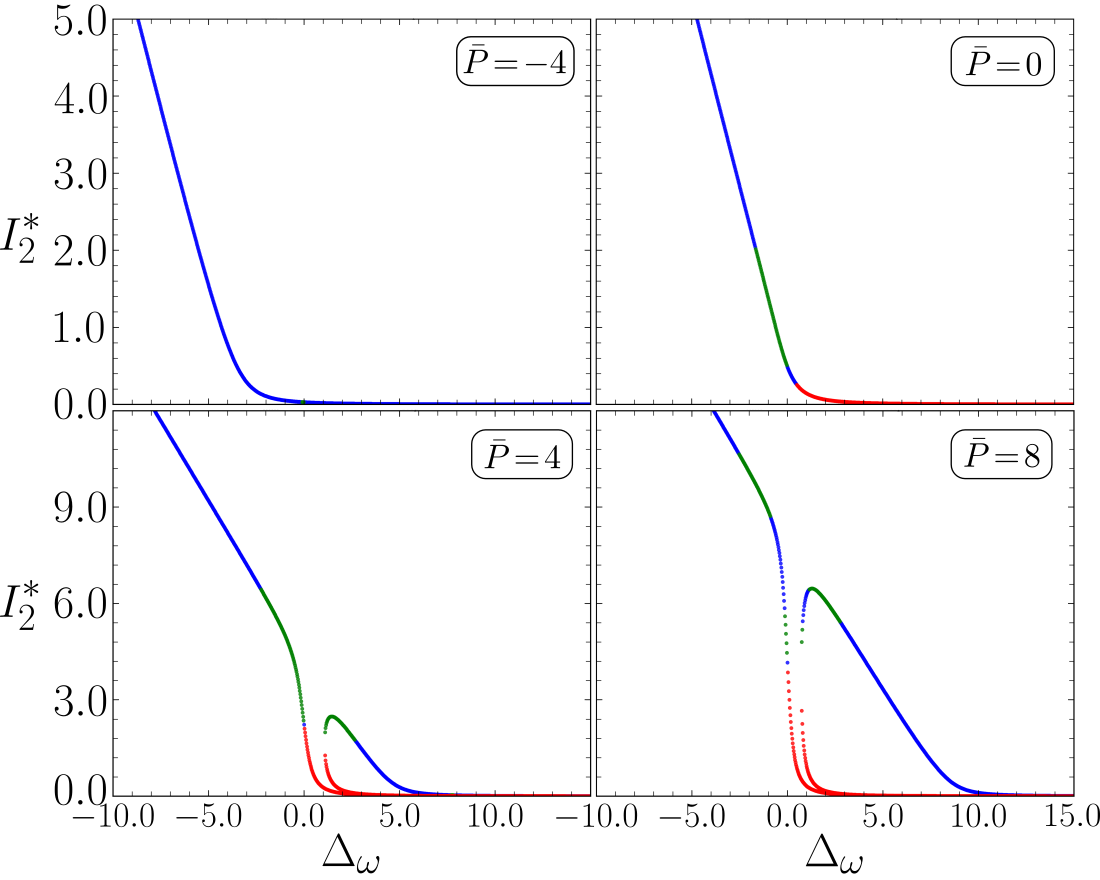}
    \caption{Bifurcation diagram for $n_1=1, n_2=0$ for different values of fixed $\bar{P}$, 
             corresponding to different horizontal lines from the left frame in figure~\ref{fig:equilibrium_wireframe}.
             Colours indicate stability, with purely unstable points in red, stable points in blue, 
             and complex unstable in green.
             $\bar{P} = -4$, $\bar{P} = 0$, $\bar{P} = 4$ and $\bar{P} = 8$ 
             correspond respectively to the magenta, blue, green and yellow curves in the top 
             frame ($\phi_2^* = 0$) of figure \ref{fig:equilibrium_wireframe}.}
\label{fig:bifurcation_map_n1=1_n2=0}
\end{figure}

\begin{figure}[h!]
    \centering
    \includegraphics[width=0.42\textwidth]{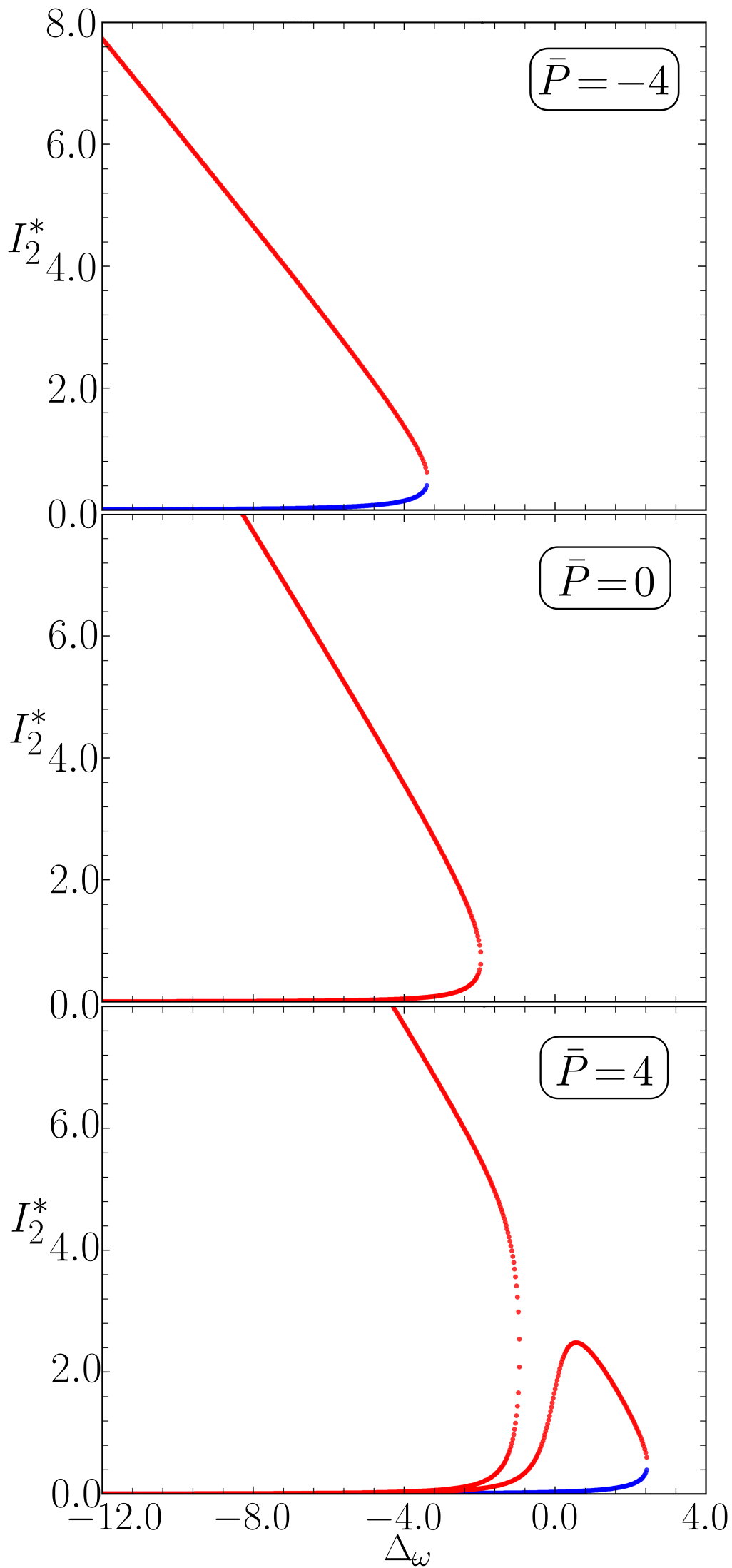}
    \caption{Bifurcation diagram for $n_1=1, n_2=1$ for different values of fixed $\bar{P}$, 
             corresponding to different horizontal lines from the right frame in figure~\ref{fig:equilibrium_wireframe}.
             Colours indicate stability, with purely unstable points in red, stable points in blue, 
             and spirally unstable in green.
             $\bar{P} = -4$, $\bar{P} = 0$ and $\bar{P} = 4$ 
             correspond respectively to the magenta, blue and green curves in the bottom 
             frame ($\phi_2^* = \pi$) of figure \ref{fig:equilibrium_wireframe}.}
\label{fig:bifurcation_map_n1=1_n2=1}
\end{figure}

In addition to the solutions themselves, the stability is promptly obtained from the 
system's jacobian, which yields a simple biquadratic form (see Appendix~\ref{sec:append:stability}). 
Figures~\ref{fig:bifurcation_map_n1=0_n2=1} to~\ref{fig:bifurcation_map_n1=1_n2=1} show 
bifurcation diagrams over the contour lines shown in figure~\ref{fig:equilibrium_wireframe}; 
in them, stable points (eigenvalues $\lambda = {\rm i} b$, for $b \in \mathbb{R}_{\neq 0}$) are 
colored blue, purely unstable points ($\lambda \in \mathbb{R}_{\neq 0}$) are colored red, 
and complex unstable ($\lambda = a + {\rm i} b$ for $a, b \in R_{\neq 0}$) 
are colored green.

When analyzing stability, we now consider the combinations of locked phases 
including $n_1$, thus resulting in four possible scenarios of ($n_1, n_2$). 
For $(n_1, n_2) = (0,0)$, all solutions are stable. 

For $(n_1, n_2) = (0,1)$, figure~\ref{fig:bifurcation_map_n1=0_n2=1} 
shows that the concave surface has its lower (upper) branch always stable (unstable). 
This indicates that at the expense of suppressing the second wave momentum, it is possible for wave 
1 to carry the particle at its minimum. 
Similarly, as the total momentum increases and the cup bifurcation takes place, the upper (lower) 
half of the branch is unstable (stable). 
For $\bar{P} > 0$, the presence of complex unstable points amidst stable ones, in both branches, 
prevents the second wave amplitude growth while the particle is locked to wave 1.

For $(n_1, n_2) = (1, 0)$, figure~\ref{fig:bifurcation_map_n1=1_n2=0} shows that higher values of 
$I_2^*$ are now stable, indicating the opposite behavior to the one found previously for 
$(n_1, n_2) = (0, 1)$. 
Now, with the increase of the second wave amplitude, the stability of its minimum allows for the 
particle to be placed at the maximum of the first wave at the same time. 
Again, as the total momentum increases, stable branches become complex unstable, but now preventing 
the locking for small $I_2^*$ amplitudes, with small stability windows, particularly for $\bar{P}=0$ 
and $\bar{P}=8$. Also, the cup-like surface provides a stable branch 
for detuning $\Delta_\omega$ higher enough. 

For $(n_1, n_2) = (1, 1)$, with the particle placed at the maxima of both waves simultaneously, 
the expected instability is found for most of the parameter space. 
However, figure~\ref{fig:bifurcation_map_n1=1_n2=1} shows that for total momentum negative enough, 
e.g. $\bar{P} = -4$, or high enough, e.g. $\bar{P} = 4$, branches with low $I_2^*$ 
are stable, with no spiral instability found. 
This indicates that at the expense of the second wave attenuation, the particle can be kept at an 
unexpected unstable position.

% Stability in the 4 scenarios:
% \begin{itemize}
    % \item $n_1 = n_2 = 0$ case is always stable (no figure for bifurcation diagram)
    % \item $n_1 = n_2 = 1$ is mostly unstable, but the bottom branch, closer to $I_2^* \approx 0 $ is stable (check this)
    % \item $n_1 = 0$, $n_2 = 1$ has two scenarios, one for negative $\bar{P}$ and other with more roots and instability change for positive $\bar{P}$.
    % \item $n_1 = 1, n_2 = 0$ has three scenarios, onf for negative $\bar{P}$, with only stability and green, other with $\bar{P}$ negative but 
    % close to zero, with instabibility, green and blue; and for $\bar{P}$ high enough, crossing the cup, with double green and all the other (blue and red)
% \end{itemize}

\begin{figure}[H]
    \includegraphics[width=0.42\textwidth]{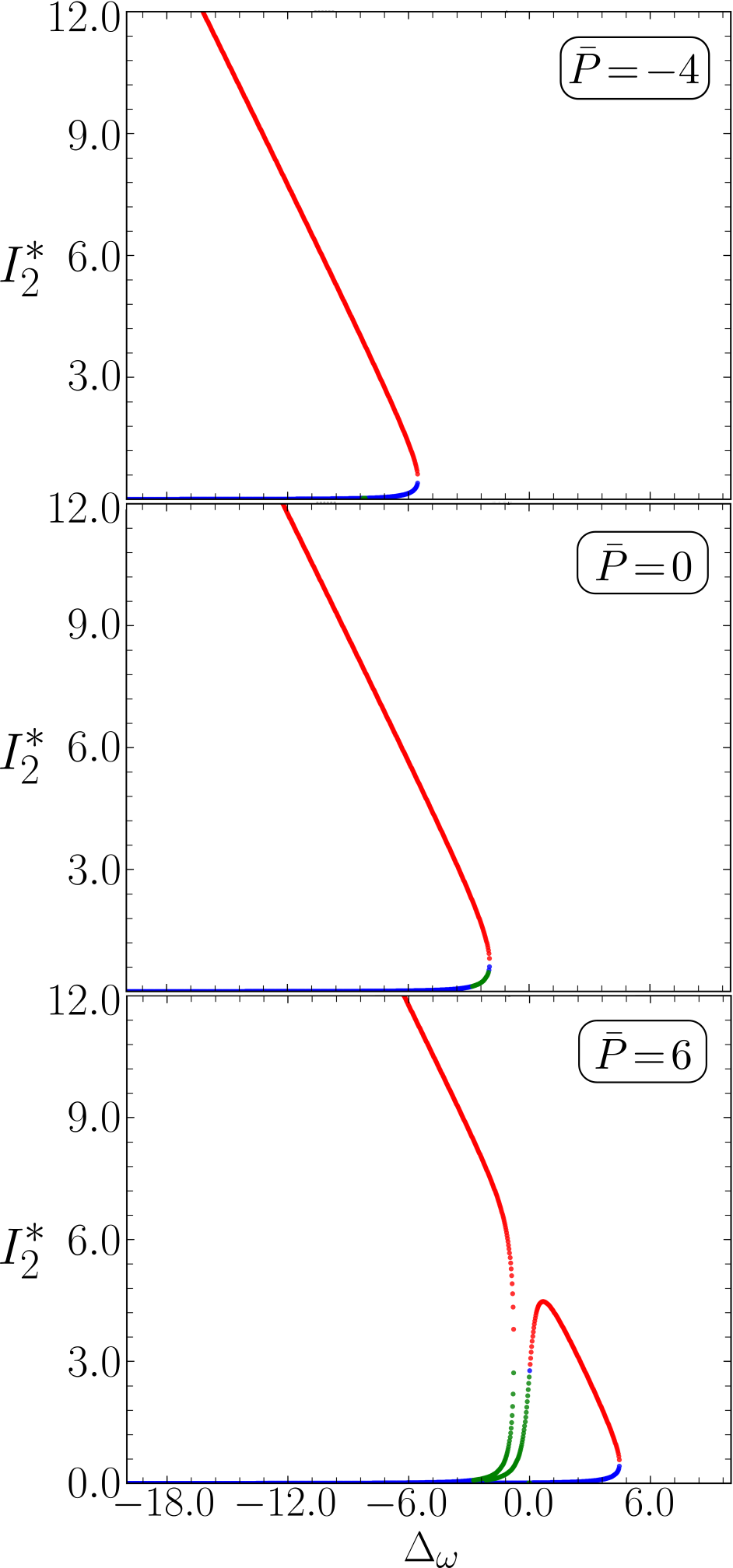}
    \caption{Bifurcation diagram for $n_1=0, n_2=1$ for different values of fixed $\bar{P}$, 
             corresponding to different horizontal lines from the right frame in figure~\ref{fig:equilibrium_wireframe}. 
             Colours indicate stability, with purely unstable points in red, stable points in blue, 
             and complex unstable in green.
             $\bar{P} = -4$, $\bar{P} = 0$ and $\bar{P} = 6$ correspond respectively to the 
             magenta, blue and orange curves in the bottom frame ($\phi_2^* = \pi$) 
             of figure \ref{fig:equilibrium_wireframe}.}
\label{fig:bifurcation_map_n1=0_n2=1}
\end{figure}

\section{Poincar\'{e} sections and chaos}\label{sec:phase-space}

Once the 3-degrees-of-freedom Hamiltonian (\ref{eq:hamiltonian-n2m1-carte}) (or in polar form 
(\ref{eq:hamiltonian-n2m1-polar})) is reduced to the equivalent 2-degree-of-freedom form 
(\ref{eq:hamiltonian-reduc-carte}) (or in polar form (\ref{eq:hamiltonian-reduc-polar})), 
trajectories can be visualized in phase space.
To this end, we made use of the Poincar\'{e} section defined as 
\begin{equation}\label{eq:pss}
    \Sigma = \lt\{\lt(\phi_1, \phi_2, I_1, I_2\rt) \in \mathbb{R}^4;\; \phi_1 = \phi_2;\; \dot{\phi_2} > 0\rt\}.
\end{equation}

% It is worth mentioning that Hamiltonian (\ref{eq:hamiltonian-reduc-carte}) is fourth degree in both 
% $u_i$ and $v_i$, making the visualization of phase space through Poincar\'{e} sections not trivial.
% Since trajectory intersections with the section may yield two positive or negative roots, 
% the usual method of fixing a position (e.g. $u_2=0$) and a direction for the surface section 
% (e.g. $v_2 > 0$), is not enough to produce a uniquely defined map. 
% Since additional roots may exist and appear over the region of the map made by the first set of roots, 
% invariant circles appear overlapped. 
% An effective solution can be achieved by filtering these two sets of roots with an additional condition, 
% such as $\dot{u}_2 > 0$.

From the selected Poincar\'{e} section, we measured the area of phase space occupied either 
by stability islands or chaotic regions as a function of the control parameters $(\bar{P}, \Delta_\omega, H)$, 
once fluctuations are expected given the nonlinear nature of the model.
For this purpose, the chaotic/regular areas were measured over the section $\Sigma$ (eq. (\ref{eq:pss})) via 
a smaller alignment index (SALI) method.
We refrain from detailing the method here and direct the reader to the original reference as presented 
by Ch. Skokos~\cite{Skokos_paper,Skokos_book}. Briefly, the algorithm integrates a single trajectory to 
numerically determine whether it is stable or chaotic, along with two deviation vectors evolved in 
tangent space according to the linearized equations of motion. 
Based on the local linearized geometry of phase space around the trajectory, the two deviation vectors 
align with each other along the unstable manifold direction in case the trajectory is chaotic.
On the other hand, when it is stable, the local geometry is a plane tangent to a torus to which 
the vectors become parallel while keeping their relative angle non null. 
Therefore, the orbit is deemed chaotic if the deviation vectors align or stable in case they do not.

\begin{figure}[h!]
    \includegraphics[width=0.5\textwidth]{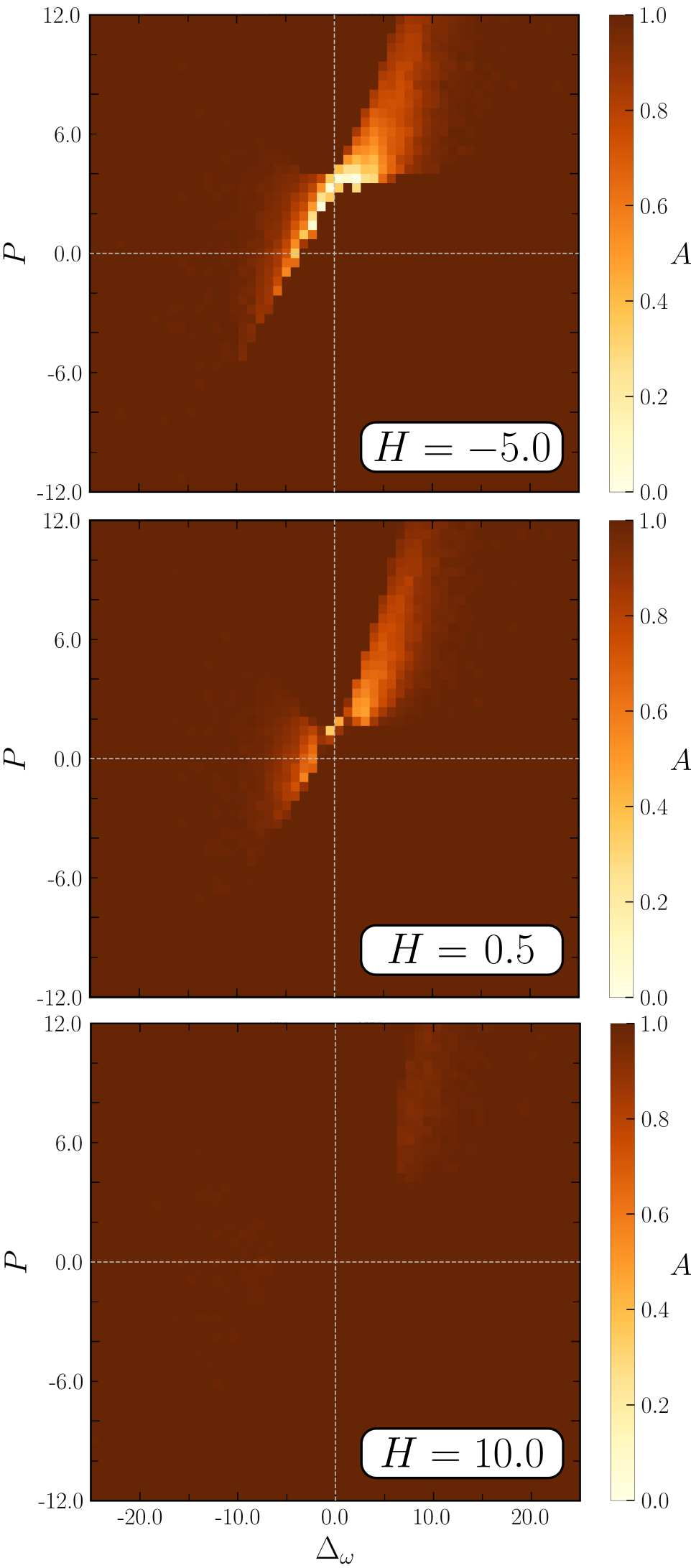}
    \caption{Color map of the stable area portion in parameter space for multiple energies $H$. 
             In the color bar, total chaos is given by $A = 0$ and total regularity by $A = 1$. 
             Grid size is $75 \times 75$.}
    \label{fig:chaotic_area}
\end{figure}

Running the SALI algorithm for each trajectory on a grid of initial conditions over the Poincar\'{e} section 
$\Sigma$ for a given set $(\bar{P}, \Delta_\omega, H)$ enables one to evaluate the percentage of stable area in 
phase space. By repeating this while scanning the parameters $(\bar{P}, \Delta_\omega)$, we obtain the profiles 
seen in figure~\ref{fig:chaotic_area}, where each panel corresponds to a different energy $H$ value. 
It is worth noting that Hamiltonian (\ref{eq:hamiltonian-reduc-carte}) defines compact energy surfaces in the 
4-dimensional phase space, thus ensuring that the fractional area $A$ is well defined.

The numerical integration of trajectories was made with an 8th-order Runge-Kutta-Dormand-Prince method 
with adaptive step, for relative and absolute precisions $\epsilon_\textrm{abs} = \epsilon_\textrm{rel} = 10^{-13}$.
In these conditions, the total energy (\ref{eq:hamiltonian-reduc-polar}) ($E(t) = H$) 
presented maximum deviations of $|\delta E| = |E(t) - E_0| = 10^{-8}$, with deviations at least one or two orders 
of magnitude smaller for smoother orbits. To avoid evaluations of inverse-square roots in equation (\ref{eq:motion-equations}), 
the integration was carried using Hamiltonian (\ref{eq:hamiltonian-reduc-carte}) (equation \ref{eq:cartesian-equations} 
in Appendix \ref{sec:append:cartesian-equations}) and converted to phasor coordinates when needed 
using (\ref{eq:phasor-carte-transform}).

For the SALI implementation, the initial deviation vectors 
$\hat{\omega}_i = (\delta u_1, \delta u_2, \delta v_1, \delta v_2)$, for $i=1,2$, 
were chosen orthogonal to each other for all initial conditions, with
\begin{equation}
\begin{split}
    \hat{\omega}_1 = (0, 1, 0, 0) \quad\textrm{and}\quad \hat{\omega}_2 = (0, 0, 0, 1).
\end{split}
\end{equation}
The deviation vectors were normalized every $\delta t = 0.5$ time units 
to prevent overflow. The numerical threshold differentiating chaotic 
from regular orbits was SALI$(t) < 10^{-10}$.

% \begin{figure}[H]
    % \centering
    % \includegraphics[width=0.5\textwidth]{figures/tmp_sali_area_stable_norm_h0,5000_parameter_space.png}
    % \caption{Color map of the STABLE area portion in parameter space for $H = 0.5$. 
            %  Total chaos is given by $A = 0$ and total regularity by $A = 1$. 
            %  Grid size is $250 \times 250$!!!. \textcolor{red}{Contour lines will show up in this figure}}
% \label{fig:chaotic_area_h0.5}
% \end{figure}
% 
% \begin{figure}[H]
    % \centering
    % \includegraphics[width=0.5\textwidth]{figures/tmp_sali_area_stable_norm_h10,0000_parameter_space.png}
    % \caption{Color map of the STABLE area portion in parameter space for $H = 10.0$. 
            %  Total chaos is given by $A = 0$ and total regularity by $A = 1$. 
            %  Grid size is $250 \times 250$!!!. \textcolor{red}{Contour lines will show up in this figure}}
% \label{fig:chaotic_area_h10.0}
% \end{figure}
% 
% \begin{figure}[H]
    % \centering
    % \includegraphics[width=0.5\textwidth]{figures/tmp_sali_area_stable_norm_h-5,0000_parameter_space.png}
    % \caption{Color map of the STABLE area portion in parameter space for $H = -5.0$. 
            %  Total chaos is given by $A = 0$ and total regularity by $A = 1$. 
            %  Grid size is $250 \times 250$!!!. \textcolor{red}{Contour lines will show up in this figure}}
% \label{fig:chaotic_area_h-5.0}
% \end{figure}

At first, the predominance of stability in the system becomes clear as, for most of the parameter space, 
chaotic area is nearly zero (dark brown regions -- $A \approx 1$).
This should hold beyond the parameter space limits chosen here 
($\bar{P}, \Delta_\omega \in [-12,12] \times [-25,25]$) as for either large $|\bar{P}|$ or $|\Delta_\omega|$, 
the equations of motion become 
\begin{equation}\label{eq:large-parameters}
\begin{split}
    \dot{I}_1    &\approx -\sqrt{2 I_1} \sin(\phi_1), \\ 
    \dot{I}_2    &\approx -\sqrt{2 I_2} \sin(\phi_2), \\ 
    \dot{\phi}_1 &\approx \pm |\bar{P}| - \frac{\cos(\phi_1)}{\sqrt{2 I_1}}, \\ 
    \dot{\phi}_2 &\approx \pm |\bar{P}| \pm |\Delta_\omega| - \frac{\cos(\phi_2)}{\sqrt{2 I_2}};
\end{split}
\end{equation}
noting that $I_1$ and $I_2$ are limited by the total energy $E=H$ in eq. (\ref{eq:hamiltonian-reduc-polar}).
If either $I_1$ or $I_2$ is small (in view of the denominator for $\dot{\phi}_i$), their amplitude 
have limited growth rate once $\phi_i$ will evolve rapidly and the average increment in $\dot{I}_1$ 
is zero or small, therefore generating a stable scenario. 
In case only $|\Delta_\omega|$ is large, $\phi_2$ and $I_2$ become regular as well, resulting 
in a regular evolution for $\phi_1$.
Therefore, for high enough values of $|\bar{P}|$ and $|\Delta_\omega|$ stability should still dominate 
phase space, as the dynamics of waves uncouples and the system reduces to the paradigmatic 
$M=N=1$ case which is analytically treatable~\cite{Adam, Gomes, Castillo-Negrete}. 
More details on the dynamics for weak coupling are given in Appendix~\ref{sec:append:decoupling}.

Despite this predominance of stability, a prominent chaotic region appears at lower values 
of total momentum and frequency detuning, which decreases for large energy and eventually vanishes
for $H \approx 10$ (figure~\ref{fig:chaotic_area}). 
Similarly to what was previously discussed, the suppression of chaos for increasing energy 
is due to the dominance of the waves amplitude energy term in Hamiltonian (\ref{eq:hamiltonian-reduc-polar}).
For high $H$, the spatial coupling terms become negligible, as their amplitude scales linearly with 
the waves amplitude whereas the `kinetic' term scales quadratically. As a result, the systems approaches 
a free wave limit.

% \textcolor{red}{
%    ((The rest of the paper will probably include an analysis on the mechanism that generate chaos.
%    From the results gathered in the past few months, the previous idea that the existence 
%    of chaotic regions was directly connected to the unstable locked-states does not seems quite right.
%    So I may think more about on how to argue on this but probably the remaining content of the 
%    paper will be a brief disccusion of poincare sections showing that chaos always emerge as separatix 
%    chaos and secondary bifurcations. Maybe an appendix mentioning the visual correlation of 
%    locked states and the chaotic regions will still be included, but not as a main result.)) 
% }

\begin{figure}[h!]
    \includegraphics[width=0.5\textwidth]{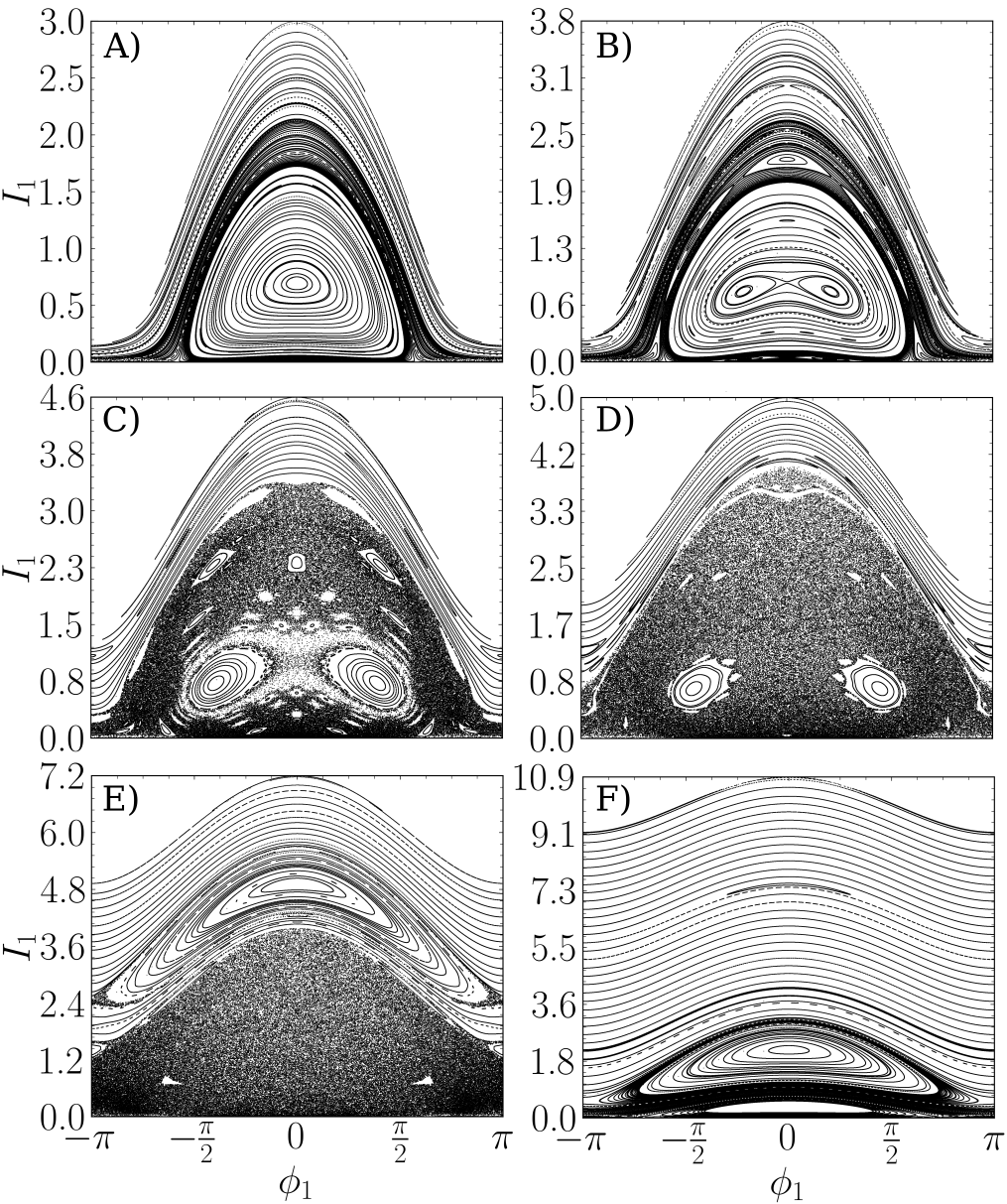}
    \caption{Phase space portraits on the Poincar\'{e} section $\Sigma$ (eq. (\ref{eq:pss})). 
             A) $\bar{P} = 0.5$, B) $\bar{P} = 1.0$, C) $\bar{P} = 1.5$, 
             D) $\bar{P} = 1.75$, E) $\bar{P} = 3.0$, F) $\bar{P} = 5.0$. 
             In all cases, $H = 0.5$ and $\Delta_\omega = 2.0$.}
\label{fig:pss_case_1}
\end{figure}

\begin{figure}[h!]
    \includegraphics[width=0.5\textwidth]{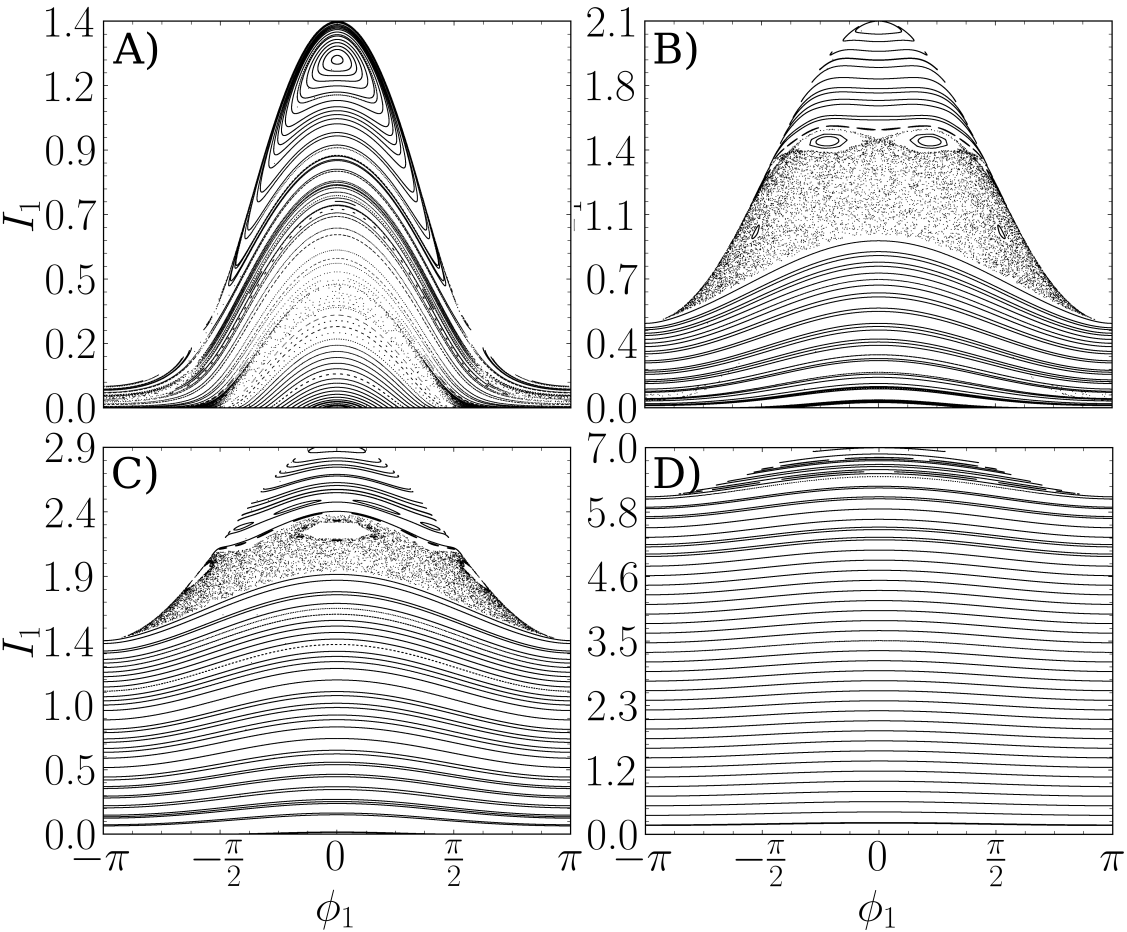}
    \caption{Phase space portraits on the Poincar\'{e} section $\Sigma$ (eq. (\ref{eq:pss})). 
             A) $\Delta_\omega = -0.5$, B) $\Delta_\omega = -4.0$, 
             C) $\Delta_\omega = -6.0$, D) $\Delta_\omega = -15.0$. 
             In all cases, $H = 0.5$ and $\bar{P} = -1.0$.}
\label{fig:pss_case_2}
\end{figure}

The emergence of these chaotic regions in phase space was seen to be of a common type, namely  
where chaos appears over and around separatrices, as illustrated by the series of portraits in 
figures~\ref{fig:pss_case_1} and~\ref{fig:pss_case_2}, for $H = 0.5$.
Figure~\ref{fig:pss_case_1} shows portraits for varying $\bar{P}$, in a parameter range 
over the first quadrant ($\bar{P}, \Delta_\omega > 0$) in figure~\ref{fig:chaotic_area}, while 
figure~\ref{fig:pss_case_2} shows portraits for varying $\Delta_\omega$ for values in the 
third quadrant ($\bar{P}, \Delta_\omega < 0$), as the system goes from regions with full regularity 
($A = 1.0$) through regions of co-existence with chaos ($A < 1.0$). 
Normally, chaos occurs within limited regions of phase space and it coexists with invariant 
circles that are often seen for higher values of amplitude $I_1$. 

In more detail, the chaotic layer as shown in figure~\ref{fig:pss_case_1} emerges 
around the main island centered at $\phi_1 = 0$ (panel A), and increases 
as secondary bifurcations disrupt the island's invariant circles. 
Simultaneously, a pitchfork bifurcation further promotes chaos within the main island 
but from the inside out (panels B, C and D). 
In panel C, it is seen that the chaotic domain, although fully connected, has a slower chaotic 
regime around the bifurcated pair of islands, with stickiness concentrated around this area.
Finally, panels E and F show that the secondary resonant islands disappear, leaving a rather uniform 
chaotic sea (panel E) that further disappears, returning to full regularity (panel F).
In a slightly different manner, the transition shown in figure~\ref{fig:pss_case_2} corresponds 
to a chaotic layer emerging in between invariant regular sets (panels B and C), although with 
the same process of secondary bifurcations taking place.

\section{Conclusions}\label{sec:conclusions}

% We show that the system can have its number of degrees of freedom reduced by simple transformations 
% and yield a simpler two degress of freedom form, keeping a good intuitive .
% By studying this simpler form, we identified
In the simple scenario of one particle interacting with two waves, although initially a 
3 degrees-of-freedom system, we show that simple canonical transformations yield a 
2 degrees-of-freedom Hamiltonian with clear connection to the original one.
On top of the transformations, the only restrictive hypotheses made were the equal 
coupling parameter between each wave and the particle, and the equal wavenumber for both 
waves.
The other assumptions made, i.e. the unit mass and unit coupling coefficients, 
can be made by rescaling variables without loss of generality.
This simple form provides a framework to understand the mechanism 
of the emergence of chaos in the system as the particle acts as a mediator of energy-momentum 
exchange between the waves, which are assumed to be non-interacting in more basic models.

For this simplified Hamiltonian, it was shown that the number of equilibrium solutions 
has a non-trivial dependence on the control parameters, namely the total momentum $\bar{P}$ 
and frequency detuning $\Delta_\omega = \omega_2 - \omega_1$. 
These equilibria correspond to any locked configuration, where the particle 
travels along with the waves while aligned with their extremum points.
For any of these spatially locked states, the nonlinear nature of the wave-particle coupling 
can yield up to 4 combinations of wave amplitudes allowing for equilibrium.
Moreover, the highest number of equilibria combinations was found for $\bar{P} > 0$, 
indicating that the waves require a minimum amount of momentum in order to carry 
the particle along with them.
This comes as a consequence of the topology of the waves amplitude equilibrium function 
as two disjoint surfaces with concavity based on the type of locked position selected.

% -- difference from other variations of the model: 
% The break of barriers as seen for the case where the electron is just a test 
% particles was not seem here

Despite the interaction between one particle and two waves being nearly integrable throughout most 
of the domain of its control parameters, the self-consistent coupling term added in our model was 
able to significantly induce chaos.
In the limit where both the total momentum $\bar{P}$ and the frequency detuning are small (up to 
order of magnitude $\bar{P} \approx 10$), the energy exchange between the waves, as mediated by the particle, is enough to 
destabilize trajectories. 
This possibility of energy-momentum exchange promoted by the non-linear coupling prevents a dynamics 
where the particle is either free or enslaved to one of the waves, as in the cases for large $|\bar{P}|$ 
or $|\Delta_\omega|$.
The emergence of chaos was seen around separatrices and higher order bifurcations, as usually observed 
in a KAM description.
However, an ergodic limit, that could be related to a destabilization of the particle motion, 
was not found in the equal wave limit tested here, which could naturally be investigated as 
an extension of the current work.

Beyond the first exploratory results shown here, the extent to which the coupling strength parameters 
$\beta_i$ can differ in order to promote enough energy-momentum exchange to produce chaos is still unclear. 
By assuming equal coupling strength, we restrict ourselves to cases where the dispersion 
relation is broad as a function of the wave-length, as in the case of a cold plasma. 
However, with different couplings, the imbalance between wave modes might suppress instabilities and therefore 
global chaos.

In the same way, a difference in wavenumbers could induce chaotic behavior, although the analysis in 
this case becomes cumbersome as the equations lose spatial periodicity and the number of free parameters 
increases.
More generally, one could evaluate how to extend the energy-momentum exchange mechanism to the limit 
of many particles and many waves, thus approaching the complete model.

\begin{acknowledgments}
We acknowledge the financial support from the scientific agencies: 
S\~{a}o Paulo Research Foundation (FAPESP) under Grant No. 2018/03211-6;
Conselho Nacional de Desenvolvimento Cient\'{i}fico e Tecnol\'{o}gico (CNPq) 
under Grants No. 200898/2022-1 and 304616/2021-4.
Coordena\c{c}\~{a}o de Aperfei\c{c}oamento de Pessoal de N\'{i}vel Superior 
(CAPES) and Comit\'{e} Fran\c{c}ais d'\'{E}valuation de
la Coop\'{e}ration Universitaire et Scientifique avec le
Br\'{e}sil (COFECUB) under Grants CAPES/COFECUB 8881.143103/2017-1 and 
COFECUB 40273QA-Ph908/18.
Centre de Calcul Intensif d'Aix-Marseille is also acknowledged for granting 
access to its high-performance computing resources.

Y. Elskens is grateful to Serge Aubry for his fascinating lectures in 1985 at Beg-Rohu, 
which revealed more facets of Hamiltonian dynamics and its broader range of applications, 
and for the enthusiasm Serge always kindly shared with his audience.
\end{acknowledgments}

\section*{Data availability}

The data that support the findings of this study are available from the corresponding author upon 
reasonable request.

\appendix

\section{Parameter reduction via scalings}\label{sec:append:param-reduc}

In the simplification step of Hamiltonian (\ref{eq:hamiltonian-full-polar}) to (\ref{eq:hamiltonian-n2m1-polar}), 
we impose equal wave number for both waves ($k_1 = k_2 = k$) as well as the coupling parameters 
with the particle ($\beta_1 = \beta_2$), allowing one to set $\epsilon \beta_1 = \epsilon \beta_2 = \beta$. 
To eliminate $k$ from within the coupling term, the position can be normalized to $ x' = k x$.
For the remaining parameters, we set the global scaling factor
\begin{equation}
    \alpha = {(\beta^2 / m)}^{1/3}
\end{equation}
and rescale variables according to
\begin{equation}
\begin{split}
    t' &= \alpha t, \\
    \omega'_i &= \alpha^{-1} \omega_i, \\
    p' / p &= I'_1 / I_1 = I'_2 / I_2 = P' / P = {(m \beta)}^{-2/3},
\end{split}
\end{equation}
and 
\begin{equation}
    H' = {(m \beta^4)}^{-1/3} H.
\end{equation}
The system in primed variables obtained from (\ref{eq:hamiltonian-full-polar}) is the reduced model 
(\ref{eq:hamiltonian-n2m1-polar}) (for $N=1$ and $M=2$). 
This scaling also acts on the action so that $S' = {(m \beta)}^{-2/3} S$. 

\section{Cartesian form of the equations of motion}\label{sec:append:cartesian-equations} 

Hamiltonian (\ref{eq:hamiltonian-reduc-carte}) yields the following equations of motion
\begin{equation}\label{eq:cartesian-equations}
\begin{split}
    \dot{u}_1 &=  \partial_{v_1} H' =  v_1 \lt(\hlf r - \bar{P}\rt), \\
    \dot{u}_2 &=  \partial_{v_2} H' =  v_2 \lt(\hlf r - \bar{P} + \Delta_\omega\rt), \\
    \dot{v}_1 &= -\partial_{u_1} H' = -u_1 \lt(\hlf r - \bar{P}\rt) + 1, \\
    \dot{v}_2 &= -\partial_{u_2} H' = -u_2 \lt(\hlf r - \bar{P} + \Delta_\omega\rt) + 1.
\end{split}
\end{equation}
where $r \coloneqq u_1^2 + v_1^2 + u_2^2 + v_2^2$.

\section{Degenerate limit of null detuning}\label{sec:append:null-detune}

In Hamiltonian (\ref{eq:hamiltonian-n2m1-carte}) or (\ref{eq:hamiltonian-n2m1-polar}), the case 
$\omega_1 = \omega_2$ is degenerate (or equivalently for $\Delta_\omega = 0$ in 
Hamiltonian (\ref{eq:hamiltonian-reduc-carte}) or (\ref{eq:hamiltonian-reduc-polar})), as both waves 
have the same wavenumber and the same phase velocity. 
This is better investigated with the change of variables
\begin{equation}
\begin{split}
    \xi_1 = \frac{X_1 + X_2}{\sqrt{2}},\quad 
    \xi_2 = \frac{X_2 - X_1}{\sqrt{2}}, \\
    \eta_1 = \frac{Y_1 + Y_2}{\sqrt{2}},\quad\textrm{and}\quad  
    \eta_2 = \frac{Y_2 - Y_1}{\sqrt{2}},
\end{split}
\end{equation}
which is canonical. Then, the Hamiltonian can be expressed as follows
\begin{equation}\label{eq:hamiltonian-null-delta}
    H_{\mathrm{sc}}^{1, 2} =
    H_0 (p, \eta_1, x, \xi_1) + H_2 \lt(\eta_2, \xi_2\rt) + \frac{\Delta_\omega}{2} \lt(\xi_1 \xi_2 + \eta_1 \eta_2\rt)
\end{equation}
with
\begin{equation}\label{eq:H0}
\begin{split}
    H_0 (p, \eta_1, x, \xi_1) &= \frac{p^2}{2}  +  \bar{\omega} \lt(\frac{\xi_ 1^2 + \eta_1^2}{2}\rt) \\
                              &\qquad\;\;       + \sqrt{2} \lt(\eta_1 \sin(x) - \xi_1 \cos(x)\rt), \\
\end{split} 
\end{equation}
and
\begin{equation}\label{eq:H2}
\begin{split}
    H_2 (\eta_2, \xi_2) &= \bar{\omega} \lt(\frac{\xi_ 2^2 + \eta_2^2}{2}\rt),
\end{split} 
\end{equation}
for $\bar\omega \coloneqq (\omega_1 + \omega_2) / 2$.

One recognizes in $H_0$ the integrable model $H_{\mathrm{sc}}^{1, 1}$ (one wave, one particle) 
with frequency $\bar{\omega}$ and coupling $\sqrt{2}$, and in $H_2$ a harmonic oscillator (with an 
eigenfrequency matching that of the wave in $H_0$). 
When $\Delta_\omega = 0$, these systems are uncoupled, and the limit $\Delta_\omega \to 0$ can be 
investigated using KAM-type theory (noting that the Hamiltonian $H_2$ is degenerate as it has no shear).

Additionally, a Galilean transformation can be used to set $\bar\omega = 0$. Then, the limit $\Delta_\omega \to 0$ 
can also lead to slow chaos~\cite{Elskens,Neishtadt_1,Neishtadt_2}.
% In this case one may notice that the Hamiltonian $H_2$ is degenerate because it has no shear.

\section{Stability of locked solutions}\label{sec:append:stability}
By assuming the locked phases solution $\phi_i^* = n_i \pi$ for $n_i = 0,1$ 
and $i=1,2$ and corresponding amplitudes $I_1^*, I_2^*$, the jacobian for 
equilibrium solutions reads 
\begin{equation*}
\mathbb{J} = \begin{pmatrix}
               0 & 0 & f_1 & 1 \\
               0 & 0 & 1 & f_2 \\
               g_1 & 0 & 0 & 0 \\
               0 & g_2 & 0 & 0 \\
             \end{pmatrix}
\end{equation*}
where $f_i$ and $g_i$ are defined as
\begin{equation*}
\begin{split}
    & f_i = f(I_i^*, n_i) \coloneqq 1 + \frac{(-1)^{n_i}}{2\sqrt{2 (I_i^*)^3}} \qquad\textrm{and} \\
    & g_i = g(I_i^*, n_i) \coloneqq -(-1)^{n_i} \sqrt{2 I_i}, \qquad\textrm{for}\qquad i=1,2.
\end{split}
\end{equation*}

Given the biquadratic form of the jacobian's characteristic polynomial 
\begin{equation*}
    \lambda^4 - \lt(g_1 f_1 + g_2 f_2\rt) \lambda^2 + g_1 g_2 (f_1 f_2 - 1) = 0,
\end{equation*}
one may solve it analytically for its eigenvalues $\lambda_i$ as
\begin{equation*}
    \lambda_{\pm,\pm} = \pm\sqrt{\Lambda_\pm},
\end{equation*}
with
\begin{equation*}
    \Lambda_\pm = \hlf \lt(g_1 f_1 + g_2 f_2 \pm \sqrt{\lt(g_1 f_1 - g_2 f_2\rt)^2 + 4 g_1 g_2}\rt),
\end{equation*}
and where all the four possible sign combinations are considered in the notation $\pm$.

\section{Decoupling limit}\label{sec:append:decoupling}

When analysing the system with weak coupling, that is in the limit $\beta = m \ll 1$, 
one recovers the paradigmatic 1.5-degree-of-freedom model of a particle slaved to two 
free waves. 
As done in appendix~\ref{sec:append:param-reduc}, we set $k_1 = k_2 = k$ and $\epsilon \beta_1 = \epsilon \beta_2 = \beta$ 
in Hamiltonian~(\ref{eq:hamiltonian-full-polar}), for the case $N=1, M=2$.
Setting $p' = p/m$ in the equations of motion then yields
\begin{equation*}
\begin{split}
    \ddot{x} &= - \sqrt{2 I_1} \sin (x - \theta_1) - \sqrt{2 I_2} \sin (x - \theta_2) \\
    \dot{I_i} &= \beta  \sqrt{2 I_i} \sin (x - \theta_i) \\
    \dot{\theta_i} - \omega_i &= \beta  {(2 I_i)}^{-1/2} \cos (x - \theta_i),
\end{split}
\end{equation*}
so that the particle evolves with the waves whereas the waves themselves evolve under an O($\beta$) feedback, 
becoming free waves in the limit $\beta \to 0$. 
This scale separation does not satisfy the standard KAM hypotheses, as the waves dynamics are degenerate (harmonic oscillators).

An alternative way to apply KAM theory to the current model would be to consider 
Hamiltonian (\ref{eq:hamiltonian-reduc-polar}) in the form
\begin{equation*}
    H' = H_1 (I'_1, \phi_1) + H_2 (I'_2, \phi_2) + I'_1 I'_2,
\end{equation*}
where
\begin{equation}
    H_1 = {I'_1}^2 / 2 - \bar P I'_1 - \sqrt{2 I'_1} \cos \phi_1
\end{equation}
and
\begin{equation}
    H_2 = {I'_2}^2 / 2 - (\bar P - \Delta_\omega) I'_2 - \sqrt{2 I'_2} \cos \phi_2,
\end{equation}
where both $H_1$ and $H_2$ reduce by a mere Galilean transformation to the $N = M = 1$ integrable model,
(eq. (9) in Gomes et al~\cite{Gomes}). We leave these analyses for future work.

\bibliographystyle{unsrt}
\bibliography{ref.bib}

\end{document}